\documentclass[12pt,a4paper,dvips]{article}
\usepackage{a4p}
\usepackage{cite,mcite,epsfig,rotating}
\usepackage{graphicx}
\usepackage{physics }
\usepackage{l3_title,ifthen, Lep}
\date{June 18, 1999}
\preprint{99-083}
\newcommand{\rar}{\rightarrow}

\def\mm{\mbox{$M \hskip 0.010in \! \! \! \! \! / _{ll}$}}

\def\Nl{\ifmmode \mathrm{N_\ell} \else
                $\mathrm{N_\ell}$   \fi}%
\def\Ne{\ifmmode \mathrm{N}_e \else
                $\mathrm{N}_e$   \fi}%
\def\Nm{\ifmmode \mathrm{N}_\mu \else
                $\mathrm{N}_\mu$   \fi}%
\def\Nt{\ifmmode \mathrm{N}_\tau \else
                $\mathrm{N}_\tau$   \fi}%
\def\Ul{\ifmmode U_{\ell N} \else
                $U_{\ell N}$   \fi}%
\def\Ue{\ifmmode U_{eN} \else
                $U_{eN}$   \fi}%
\def\MZ{m_{\mathrm{Z}}}
\def\MW{m_{\mathrm{W}}}
\def\MN{m_{\mathrm{N}}}
\def\MVI{m_{\mathrm{\it vis}}}

\def\Ne{\ifmmode \mathrm{N}_\mathrm{e}\else
                $\mathrm{N}_e$\fi}%
\def\Nm{\ifmmode \mathrm{N}_\mu\else
                $\mathrm{N}_\mu$\fi}%
\def\Nt{\ifmmode \mathrm{N}_\tau\else
                $\mathrm{N}_\tau$\fi}%

\newlength{\capindent}
\newlength{\capwidth}
\newlength{\figwidth}
\newcommand{\icaption}[2][!*!,!]{\hspace*{\capindent}%
  \begin{minipage}{\capwidth}
    \ifthenelse{\equal{#1}{!*!,!}}%
      {\caption{#2}}%
      {\caption[#1]{#2}}
  \end{minipage}}
\begin{document}
\bibliographystyle{l3style}
\begin{titlepage}
 \title{ Search for Heavy Isosinglet Neutrinos in e$^+$e$^-$ Annihilation
at 130 {\boldmath $< \sqrt{s} <$ } 189 {\boldmath $\GeV$}  }
\author{The L3 Collaboration}
%
%
\begin{abstract}

A search for heavy neutrinos that are isosinglets
under the standard $SU(2)_L$ gauge group is made at center-of-mass energies
130 $< \sqrt{s} <$  189 \GeV\ with the L3 detector at LEP.
Such heavy  neutrinos 
are expected in many extensions of the Standard Model.
The search is performed for the first generation heavy singlet neutrino, $\Ne$,
through the decay mode $\Ne \rightarrow \mathrm{e} + \mathrm{W}$.   
We set  upper 
limits on the mixing parameter between the heavy and light neutrino
for the heavy neutrino
mass range from 80 \GeV\ to 185 \GeV.
   
\end{abstract}

%
%
\submitted
 \end{titlepage}

\section*{Introduction}

In the Standard Model all fundamental fermions have a right-handed 
component that transforms as an isosinglet 
except the neutrinos, which are observed only in left-handed form.
However, isosinglet heavy neutrinos
arise in many models that attempt to unify the presently known interactions into 
a single gauge scheme, such as Grand Unified Theories or Superstring inspired
models \cite{Valle2}. Their existence is also predicted in many extended 
electroweak models such as left-right symmetric and see-saw models \cite{seesaw}. 


In e$^+ $e$^-$ annihilation isosinglet heavy neutrinos can be produced through their 
mixing with the light neutrinos (see Figure~1). Constraints on 
 isosinglet neutrino
mixing have been set by several experiments~\cite{otherINHL,lepINHL}. Heavy
singlet neutrinos have been searched for in leptonic decays of mesons and in neutrino
beam experiments~\cite{otherINHL}, leading to stringent upper limits on 
the square of
the mixing amplitude, $|U_\ell|^2$, reaching 10$^{-7}$ in the low mass 
region ($\MN$ below
2~to~3~\GeV). In addition, 
three LEP experiments~\cite{lepINHL} have 
 set
limits on $|U_\ell|^2$ of the order of 
$10^{-3}$~to~$10^{-5}$ for the neutrino mass range from 
3 \GeV\ up to 80 \GeV.

In this paper we report on a direct search for isosinglet heavy neutrinos with
masses larger than the W mass.

\section*{Production and decay}

In this search, one isosinglet neutral heavy lepton \Nl is assumed to be
associated with each generation of light neutrinos via the mixing amplitude
$U_\ell$. We do not consider
mixing of the light neutrinos with higher isodoublet states
(sequential leptons) nor the possibility of mixing among light neutrinos
(as discussed in Ref. \cite{Gronau}).

The mixing between the isosinglet neutral lepton and its associated
isodoublet neutrino allows single production to occur in e$^+ $e$^-$ 
annihilation:
\vskip 0.5cm
\centerline{e$^+ $e$^- \rightarrow \Nl + \nu_{\ell} $.}            
\vskip 0.5cm  
In contrast to sequential isodoublet neutral leptons where  
pair production is dominant (when kinematically allowed),
here single production dominates. The 
corresponding pair production cross section is suppressed
relatively to the single production cross section by 
an additional $|U_\ell|^2$ factor, which is expected to be small
(from indirect searches at LEP1 and low-energy experiments 
$|U_{\ell}|$ $\leq$ 0.1 for $\MN$ larger than
 80~\GeV~\cite{singlet_lim}).
The single production process proceeds through $s$-channel Z exchange for
all generations. For
the first generation, $\Ne$, when heavy neutrinos can couple to
electrons, there is an additional contribution from $t$-channel W exchange
(see Figure~1).
Due to this additional contribution the production cross section for the
first generation heavy neutrinos  can be as
high as 0.6 pb.
Since the mixing amplitude is expected
 to be small,
the production cross section for~\Nm~and~\Nt~is too small to
be explored at LEP2. Therefore, in this paper we concentrate on the search 
for the 
first generation heavy singlet neutrino, $\Ne$. 

Isosinglet heavy neutrinos decay via the neutral or charged weak currents
through the mixing with a light lepton:
\[ \Ne
 \rightarrow \mathrm{Z}\nu_\mathrm{e} \mbox{~or~}
 \Ne \rightarrow \mathrm{We}. \]

%

For neutrino masses not far above the W mass, heavy neutrinos
decay predominantly through the W boson due to  
phase space suppression of the decay into Z, while for large masses
$Br(\Ne \rightarrow  \mathrm{eW}) = 0.67$ and
$Br(\Ne \rightarrow \mathrm{Z}\nu_\mathrm{e} ) = 0.33$~\cite{singlet_djo}.

\section*{Data sample and event simulation}

We present the analysis of data collected by L3 \cite{hlep04} 
at LEP2 from 1995 to 1998. The search was performed using a luminosity of
12.1 pb$^{-1}$ at $\sqrt{s} = 130$~to~$136$ \GeV, 
21.1 pb$^{-1}$ at $\sqrt{s} = 161$~to~$172$ \GeV,
55.2 pb$^{-1}$ at $\sqrt{s}$=183 \GeV, and 176 pb$^{-1}$ at 
$\sqrt{s}$=189 \GeV. 

Using  the full differential cross section~\cite{singlet_buh},
 a dedicated Monte Carlo
generator is constructed to simulate heavy 
singlet neutrino production and decay.
 Subsequent hadronic fragmentation and decay are
simulated by the JETSET Monte Carlo program~\cite{hlep07}. Initial and final
state radiation is taken into account. In addition, we include the effects
of the finite width of the produced W and Z bosons.
For the search we considered the 
mass range of the heavy neutrino between 80~and~185~\GeV.
For the simulation of background from Standard Model processes, the
following Monte Carlo programs are used:
PYTHIA~5.7~\cite{hlep07} ($\mbox{e}^+ \mbox{e}^- \rar q \bar q (\gamma),
~\mbox{Z} \mbox{e}^+ \mbox{e}^-,~\mbox{Z} \mbox{Z}$),
KORALZ~\cite{hlep071} ($\mbox{e}^+ \mbox{e}^- \rar \tau^+ \tau^- (\gamma)$),
KORALW~\cite{hlep072} ($\mbox{e}^+ \mbox{e}^- \rar \mbox{W}^+ \mbox{W}^- $),
PHOJET~\cite{hlep08} ($\mbox{e}^+ \mbox{e}^- \rar \mbox{e}^+ \mbox{e}^- \mbox{q} \bar{\mbox{q}}$),
DIAG36~\cite{hlep081} ($\mbox{e}^+ \mbox{e}^- \rar \mbox{e}^+ \mbox{e}^- \tau^+ \tau^-$),
and EXCALIBUR~\cite{hlep082} ($\mbox{e}^+ \mbox{e}^- \rar \mbox{f} \bar{\mbox{f}}'
\mbox{f} \bar{\mbox{f}}'$).

The Monte Carlo events have been simulated in the L3 detector
using the GEANT3 program~\cite{hlep09}, which takes into account the
effects of energy loss, multiple scattering and showering in the
materials.

\section*{Event signatures and selection}

The most important backgrounds for heavy neutrino searches are 
W$^+$W$^-$production  
with one hadronic and one leptonic W decay (96\% of the background),
$q \bar q (\gamma)$ (3.6\%) and ZZ production (0.4\%). 
Reducing the W$^+$W$^-$ background requires full reconstruction
of the heavy neutrino mass from its decay products. 
The only decay channel for which this is
possible is $ \Ne \rightarrow\mathrm{eW}$
with $\mathrm{W} \rightarrow \mathrm{jets}$. 
Thus, the event signature is one isolated electron plus hadronic jets.
This channel has the largest branching ratio varying
between 68\% and 45\% depending on the heavy neutrino mass.

An electron is defined as a cluster in the electromagnetic 
calorimeter with an energy larger than 4 \GeV\ matched to a track 
in the $(R,\phi)$ plane to within 10 mrad. 
The cluster shower profile should be consistent with the one expected for an 
electron, {\it i.e.}
0.97~$<$~$E_9/E_{25}$~$<$~1.03, where $E_{9}$ and 
$E_{25}$ are the 
sums of the  lateral-energy-leakage-corrected
energies of 9 and 25 BGO crystals centered on the most energetic one.
The electron polar angle $\theta$ must be in the fiducial volume defined by
$|\cos \theta|<0.94$.
The energy, excluding the electron energy, deposited in a 10$^{\circ}$ cone
around the electron direction, is required to be smaller than 5 \GeV.  

Jets are reconstructed from electromagnetic and hadronic calorimeter 
clusters using the Durham algorithm \cite{singlet_jet} with a
jet resolution parameter of $y_{cut}$ = 0.008. 
The jet momenta are defined
by the vectorial energy sum of calorimetric clusters.

The event selection requires at least two hadronic jets plus one isolated 
electron.
The visible energy must be greater than 70 \GeV\ and the number of
reconstructed tracks must be greater than 6. 
The polar angle $\theta$ of the missing momentum should be in the range
$25^{\circ}<\theta<155^{\circ}$. 
The visible 
invariant mass $\MVI$ of the event is reconstructed and, to improve the
resolution, it is rescaled according to \\
$$\MN=\MVI {\sqrt s\over p_\nu+E}~,$$
where $p_\nu$
is the missing momentum of the event, and $E$ is the
visible energy. 
Figure~2 shows the rescaled invariant
mass $\MN$ of the events after all previous cuts have been 
applied. 
Two regions are defined. In region 1, where the heavy neutrino mass is close
to the W mass, a significant fraction of W's produced in
$\Ne $ decays are off-shell. In region 2,
$\MN$ $>$ 100 \GeV, the W's are produced on-shell.
In this case to
further improve the resolution on the mass measurement, the determination
of jet energies and angles, and the missing momentum direction (both for the signal
and the W$^+$W$^-$ background), a kinematic fit is applied imposing 
 four-momentum 
conservation and the constraint that the invariant mass of the hadronic jets is
equal to the W mass. 
In total, 21 events are selected in region~1, 
while 26.2 are expected from the background.
The corresponding numbers for region 2 are 464 and 463.3.
Figure~3 shows the distribution of the invariant mass of the electron and
missing momentum, $m_{\mathrm{e} \nu}$, for events in region~2 after the kinematic 
fit has been applied. One can see a clear peak at 80 \GeV\ coming from 
the W$^+$W$^-$  background. 

The final selection requires the invariant mass of the electron and missing 
momentum to be outside the W~mass region.
Applying the cut,
$m_{\mathrm{e} \nu}$~$<$~70~\GeV\ or $m_{\mathrm{e} \nu}$~$>$~90~\GeV, rejects 70\% of the 
background events.
Figure~4 shows the invariant mass of the events accepted after the
mass cut is applied.
We observe a 
good agreement between data and expected background: 84
events pass the selection, while 88.7 are expected from the SM background.

\section*{Results}

We calculate the 95\% confidence level upper limit on the square of the
mixing amplitude, $|U_{\mathrm{e}}|^2$, following the procedure in reference~\cite{singlet_obr}. 
In region 1 we use the total number of events found in data  and 
MC background  to set a limit. 
In region 2 the number of events
in data and MC background for a given mass $\MN$ is defined as the number of
events which have a reconstructed mass in the range of $\MN \pm$ 2$\sigma$,
where mass resolution $\sigma$ varies from~2~to~2.5~GeV over the mass range
considered.
The selection efficiency varies from 20\% to 45\% depending on
the heavy neutrino mass and the center-of-mass energy.
 The systematic error, which is mainly due to the uncertainty
in the energy calibration, the simulation and reconstruction 
 of the heavy neutrinos, and the Monte Carlo statistics,
 is estimated to be 5\% relative.
To obtain limits, the selection efficiency has been reduced by one
standard deviation of the total systematic error.
 Taking into account
the luminosities, the selection efficiencies, the production
cross sections and branching ratio $Br(\Ne \rightarrow \mathrm{eW})$ for heavy 
singlet neutrinos at $\sqrt{s}$ = 133 to 189 \GeV,
we obtain an upper limit on the square of the mixing parameter, $|U_{\mathrm{e}}|^2$.

 The results for the mixing amplitude, $|U_{\mathrm{e}}|^2$, as a function
of the mass are shown in Figure~5. These limits are the first 
results for masses of singlet heavy neutrinos 
greater than 80 \GeV.

\section*{Acknowledgements}

We wish to express our gratitude to the CERN accelerator divisions for the 
excellent performance of the LEP machine. We acknowledge with appreciation 
the effort of the engineers, technicians and support staff who have
participated in the construction and maintenance of this experiment.

%
%
\newpage
\section*{Author List}
\typeout{   }     
\typeout{Using author list for paper 178 -?}
\typeout{$Modified: Fri Jun  4 11:46:37 1999 by clare $}
\typeout{!!!!  This should only be used with document option a4p!!!!}
\typeout{   }
%
%
%
%
%
%

\newcount\tutecount  \tutecount=0
\def\tutenum#1{\global\advance\tutecount by 1 \xdef#1{\the\tutecount}}
\def\tute#1{$^{#1}$}
\tutenum\aachen            
\tutenum\nikhef            
\tutenum\mich              
\tutenum\lapp              
\tutenum\basel             
\tutenum\lsu               
\tutenum\beijing           
\tutenum\berlin            
\tutenum\bologna           
\tutenum\tata              
\tutenum\ne                
\tutenum\bucharest         
\tutenum\budapest          
\tutenum\mit               
\tutenum\florence          
\tutenum\cern              
\tutenum\wl                
\tutenum\geneva            
\tutenum\hefei             
\tutenum\seft              
\tutenum\lausanne          
\tutenum\lecce             
\tutenum\lyon              
\tutenum\madrid            
\tutenum\milan             
\tutenum\moscow            
\tutenum\naples            
\tutenum\cyprus            
\tutenum\nymegen           
\tutenum\caltech           
\tutenum\perugia           
\tutenum\cmu               
\tutenum\prince            
\tutenum\rome              
\tutenum\peters            
\tutenum\salerno           
\tutenum\ucsd              
\tutenum\santiago          
\tutenum\sofia             
\tutenum\korea             
\tutenum\alabama           
\tutenum\utrecht           
\tutenum\purdue            
\tutenum\psinst            
\tutenum\zeuthen           
\tutenum\eth               
\tutenum\hamburg           
\tutenum\taiwan            
\tutenum\tsinghua          
{
\parskip=0pt
\noindent
{\bf The L3 Collaboration:}
\ifx\selectfont\undefined
 \baselineskip=10.8pt
 \baselineskip\baselinestretch\baselineskip
 \normalbaselineskip\baselineskip
 \ixpt
\else
 \fontsize{9}{10.8pt}\selectfont
\fi
\medskip
\tolerance=10000
\hbadness=5000
\raggedright
\hsize=162truemm\hoffset=0mm
\def\r{\rlap,}
\noindent

M.Acciarri\r\tute\milan\
P.Achard\r\tute\geneva\ 
O.Adriani\r\tute{\florence}\ 
M.Aguilar-Benitez\r\tute\madrid\ 
J.Alcaraz\r\tute\madrid\ 
G.Alemanni\r\tute\lausanne\
J.Allaby\r\tute\cern\
A.Aloisio\r\tute\naples\ 
M.G.Alviggi\r\tute\naples\
G.Ambrosi\r\tute\geneva\
% ...
\newpage
\begin{list}{A}{\itemsep=0pt plus 0pt minus 0pt\parsep=0pt plus 0pt minus 0pt
                \topsep=0pt plus 0pt minus 0pt}
\item[\aachen]
 I. Physikalisches Institut, RWTH, D-52056 Aachen, FRG$^{\S}$\\
 III. Physikalisches Institut, RWTH, D-52056 Aachen, FRG$^{\S}$
\item[\nikhef] National Institute for High Energy Physics, NIKHEF, 
     and University of Amsterdam, NL-1009 DB Amsterdam, The Netherlands
\item[\mich] University of Michigan, Ann Arbor, MI 48109, USA
\item[\lapp] Laboratoire d'Annecy-le-Vieux de Physique des Particules, 
     LAPP,IN2P3-CNRS, BP 110, F-74941 Annecy-le-Vieux CEDEX, France
\item[\basel] Institute of Physics, University of Basel, CH-4056 Basel,
     Switzerland
\item[\lsu] Louisiana State University, Baton Rouge, LA 70803, USA
\item[\beijing] Institute of High Energy Physics, IHEP, 
  100039 Beijing, China$^{\triangle}$ 
\item[\berlin] Humboldt University, D-10099 Berlin, FRG$^{\S}$
\item[\bologna] University of Bologna and INFN-Sezione di Bologna, 
     I-40126 Bologna, Italy
\item[\tata] Tata Institute of Fundamental Research, Bombay 400 005, India
\item[\ne] Northeastern University, Boston, MA 02115, USA
\item[\bucharest] Institute of Atomic Physics and University of Bucharest,
     R-76900 Bucharest, Romania
\item[\budapest] Central Research Institute for Physics of the 
     Hungarian Academy of Sciences, H-1525 Budapest 114, Hungary$^{\ddag}$
\item[\mit] Massachusetts Institute of Technology, Cambridge, MA 02139, USA
\item[\florence] INFN Sezione di Firenze and University of Florence, 
     I-50125 Florence, Italy
\item[\cern] European Laboratory for Particle Physics, CERN, 
     CH-1211 Geneva 23, Switzerland
\item[\wl] World Laboratory, FBLJA  Project, CH-1211 Geneva 23, Switzerland
\item[\geneva] University of Geneva, CH-1211 Geneva 4, Switzerland
\item[\hefei] Chinese University of Science and Technology, USTC,
      Hefei, Anhui 230 029, China$^{\triangle}$
\item[\seft] SEFT, Research Institute for High Energy Physics, P.O. Box 9,
      SF-00014 Helsinki, Finland
\item[\lausanne] University of Lausanne, CH-1015 Lausanne, Switzerland
\item[\lecce] INFN-Sezione di Lecce and Universit\'a Degli Studi di Lecce,
     I-73100 Lecce, Italy
\item[\lyon] Institut de Physique Nucl\'eaire de Lyon, 
     IN2P3-CNRS,Universit\'e Claude Bernard, 
     F-69622 Villeurbanne, France
\item[\madrid] Centro de Investigaciones Energ{\'e}ticas, 
     Medioambientales y Tecnolog{\'\i}cas, CIEMAT, E-28040 Madrid,
     Spain${\flat}$ 
\item[\milan] INFN-Sezione di Milano, I-20133 Milan, Italy
\item[\moscow] Institute of Theoretical and Experimental Physics, ITEP, 
     Moscow, Russia
\item[\naples] INFN-Sezione di Napoli and University of Naples, 
     I-80125 Naples, Italy
\item[\cyprus] Department of Natural Sciences, University of Cyprus,
     Nicosia, Cyprus
\item[\nymegen] University of Nijmegen and NIKHEF, 
     NL-6525 ED Nijmegen, The Netherlands
\item[\caltech] California Institute of Technology, Pasadena, CA 91125, USA
\item[\perugia] INFN-Sezione di Perugia and Universit\'a Degli 
     Studi di Perugia, I-06100 Perugia, Italy   
\item[\cmu] Carnegie Mellon University, Pittsburgh, PA 15213, USA
\item[\prince] Princeton University, Princeton, NJ 08544, USA
\item[\rome] INFN-Sezione di Roma and University of Rome, ``La Sapienza",
     I-00185 Rome, Italy
\item[\peters] Nuclear Physics Institute, St. Petersburg, Russia
\item[\salerno] University and INFN, Salerno, I-84100 Salerno, Italy
\item[\ucsd] University of California, San Diego, CA 92093, USA
\item[\santiago] Dept. de Fisica de Particulas Elementales, Univ. de Santiago,
     E-15706 Santiago de Compostela, Spain
\item[\sofia] Bulgarian Academy of Sciences, Central Lab.~of 
     Mechatronics and Instrumentation, BU-1113 Sofia, Bulgaria
\item[\korea] Center for High Energy Physics, Adv.~Inst.~of Sciences
     and Technology, 305-701 Taejon,~Republic~of~{Korea}
\item[\alabama] University of Alabama, Tuscaloosa, AL 35486, USA
\item[\utrecht] Utrecht University and NIKHEF, NL-3584 CB Utrecht, 
     The Netherlands
\item[\purdue] Purdue University, West Lafayette, IN 47907, USA
\item[\psinst] Paul Scherrer Institut, PSI, CH-5232 Villigen, Switzerland
\item[\zeuthen] DESY-Institut f\"ur Hochenergiephysik, D-15738 Zeuthen, 
     FRG
\item[\eth] Eidgen\"ossische Technische Hochschule, ETH Z\"urich,
     CH-8093 Z\"urich, Switzerland
\item[\hamburg] University of Hamburg, D-22761 Hamburg, FRG
\item[\taiwan] National Central University, Chung-Li, Taiwan, China
\item[\tsinghua] Department of Physics, National Tsing Hua University,
      Taiwan, China
\item[\S]  Supported by the German Bundesministerium 
        f\"ur Bildung, Wissenschaft, Forschung und Technologie
\item[\ddag] Supported by the Hungarian OTKA fund under contract
numbers T019181, F023259 and T024011.
\item[\P] Also supported by the Hungarian OTKA fund under contract
  numbers T22238 and T026178.
\item[$\flat$] Supported also by the Comisi\'on Interministerial de Ciencia y 
        Tecnolog{\'\i}a.
\item[$\sharp$] Also supported by CONICET and Universidad Nacional de La Plata,
        CC 67, 1900 La Plata, Argentina.
\item[$\natural$] Supported by Deutscher Akademischer Austauschdienst.
\item[$\diamondsuit$] Also supported by Panjab University, Chandigarh-160014, 
        India.
\item[$\triangle$] Supported by the National Natural Science
  Foundation of China.
\item[\dag] Deceased.
\end{list}
}
\vfill





\newpage


\newpage




\begin{figure}[p]
\begin{center}
\mbox{\epsfysize=16cm\epsffile{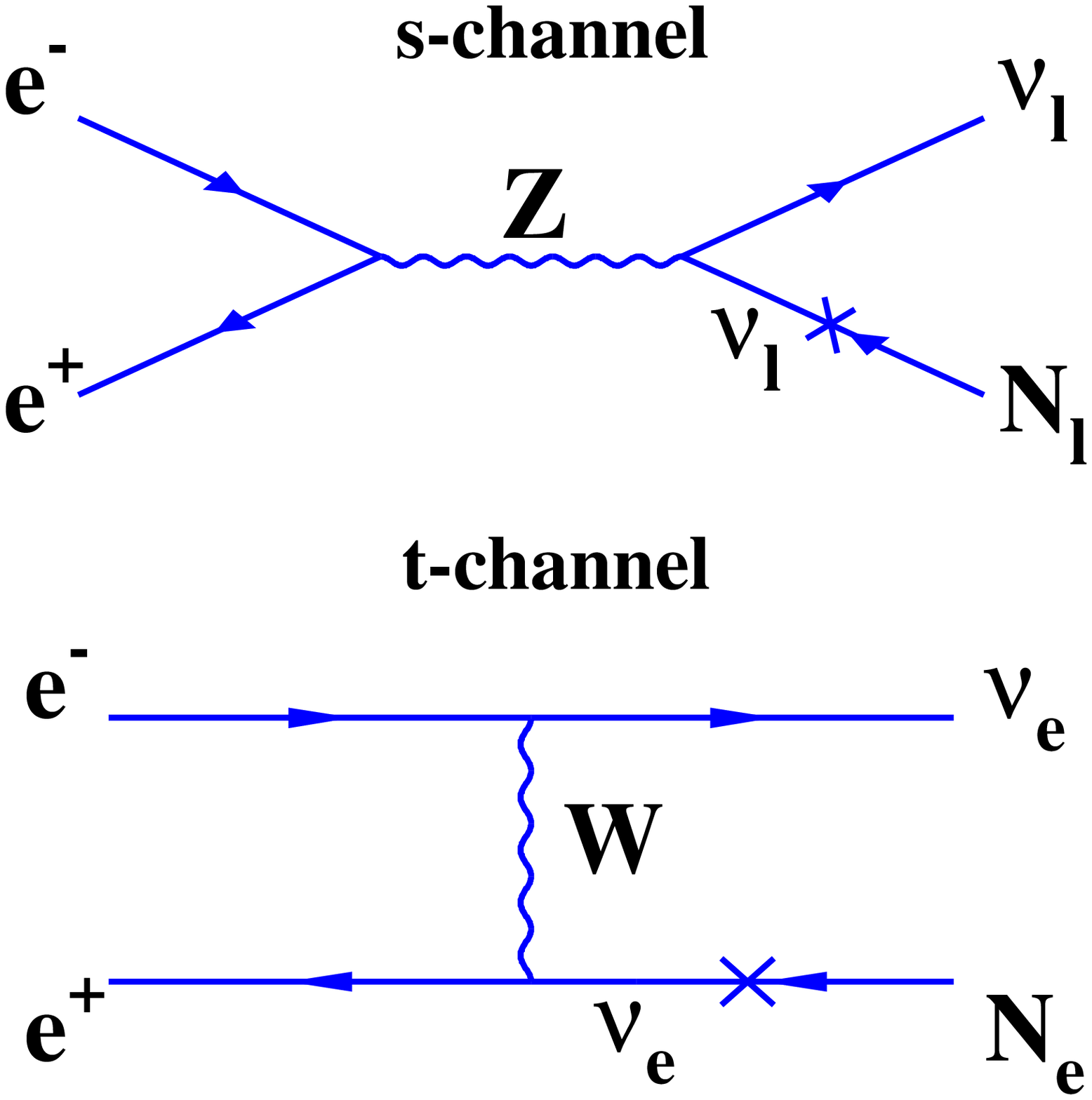}}
\end{center}
\caption
{
Feynman diagrams showing the production of isosinglet heavy neutrinos.
Here the lepton $\ell$ denotes $\mathrm{e}$, $\mu$, or $\tau$ for
$s$-channel production.
}
\label {bla}
\end{figure}

\begin{figure}[p]
\begin{center}
\mbox{\epsfysize=16cm\epsffile{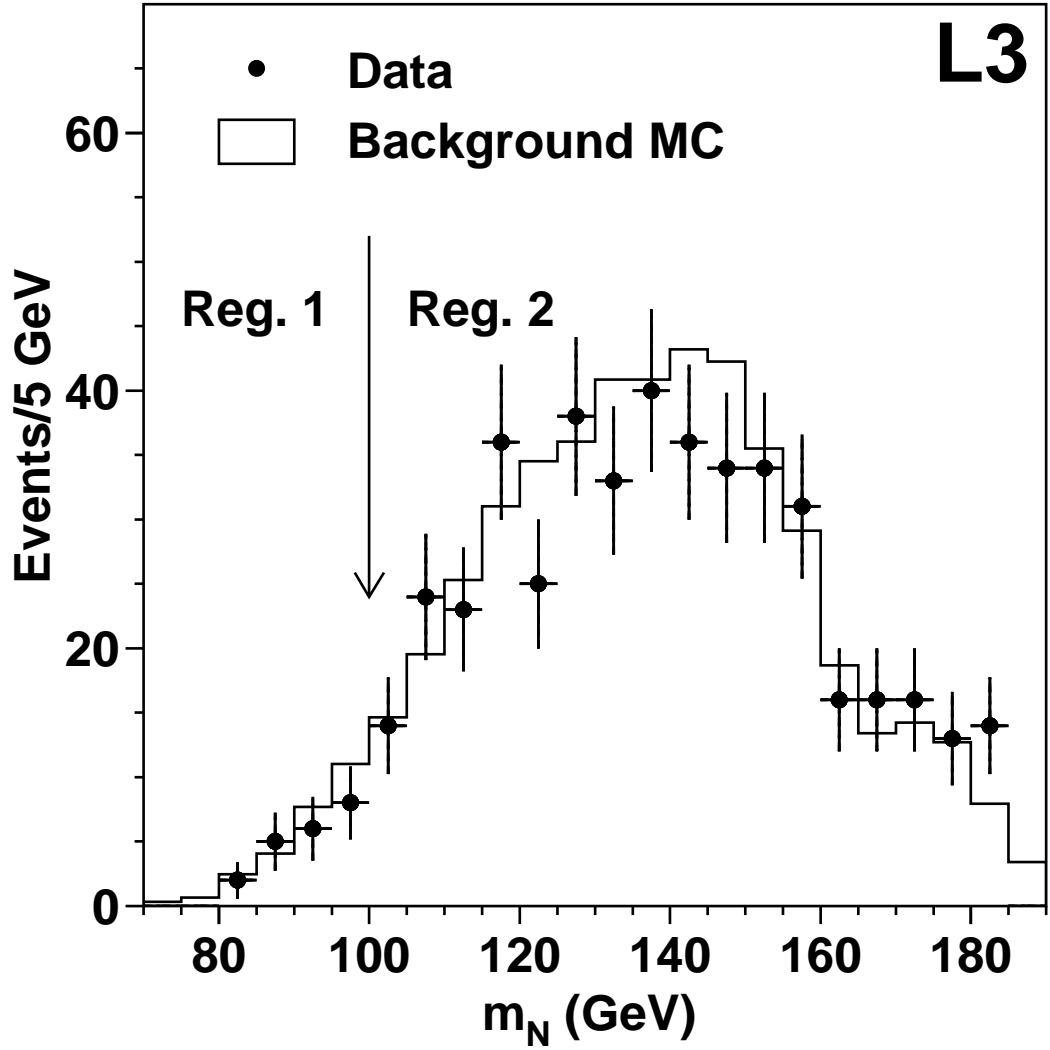}}
\end{center}
\caption
{
Distribution of the rescaled visible invariant mass, $\MN$, of the event. 
The points are the
data and the solid histogram is the background. See text for definitions
of Region 1 and Region 2.
}
\label {fig3}
\end{figure}

\begin{figure}[p]
\begin{center}
\mbox{\epsfysize=16cm\epsffile{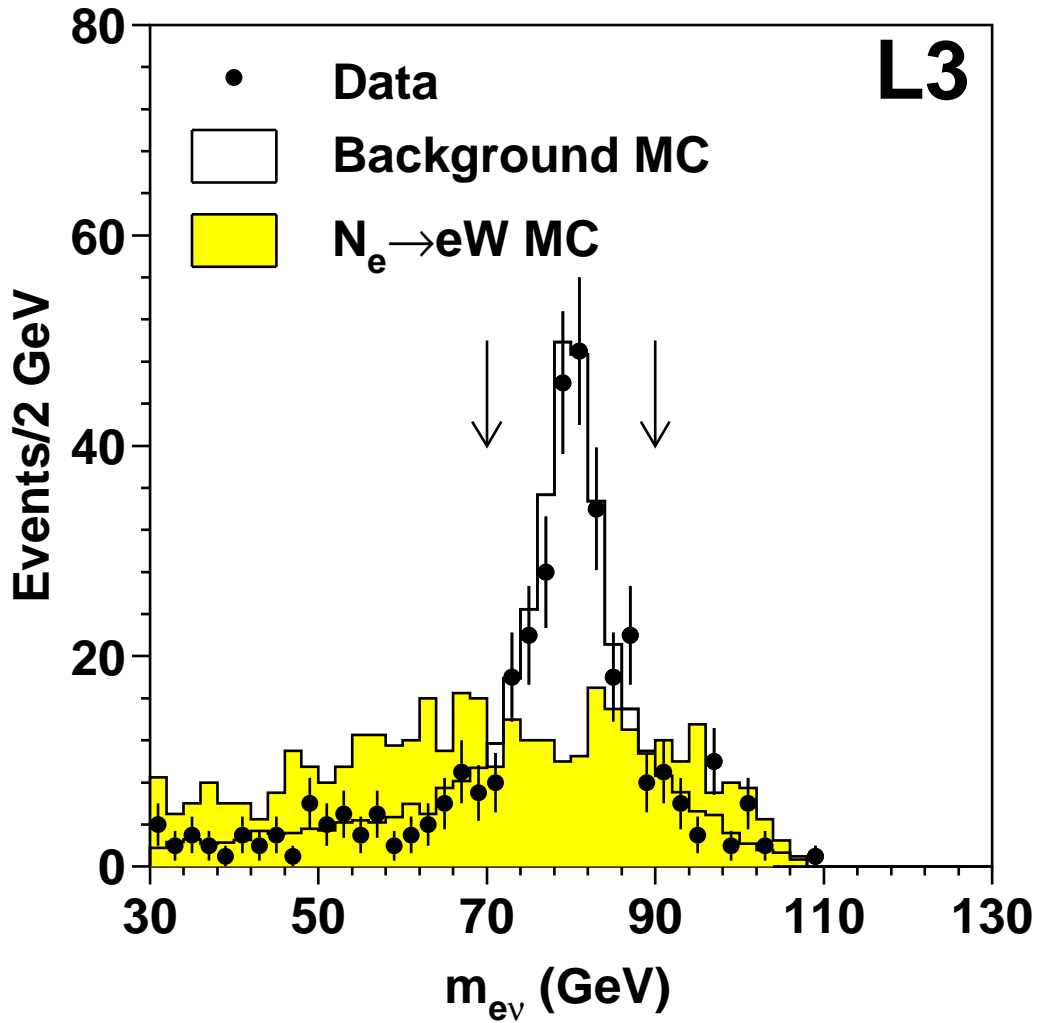}}
\end{center}
\caption
{
The invariant mass, $m_{\mathrm{e} \nu}$, of the isolated electron and missing momentum.
The points are the data taken at $\sqrt{s}$ = 189 \GeV, the solid histogram
is the background Monte Carlo. The shaded histogram is the predicted
signal e$^+ $e$^- \rightarrow \nu \mathrm{N}$ for a 140 \GeV\ heavy
neutrino. 
The normalization for the signal Monte Carlo is arbitrary. 
The arrows indicate the value of the applied
cut.
}
\label {ainwl}
\end{figure}

\begin{figure}[p]
\begin{center}
\mbox{\epsfysize=16cm\epsffile{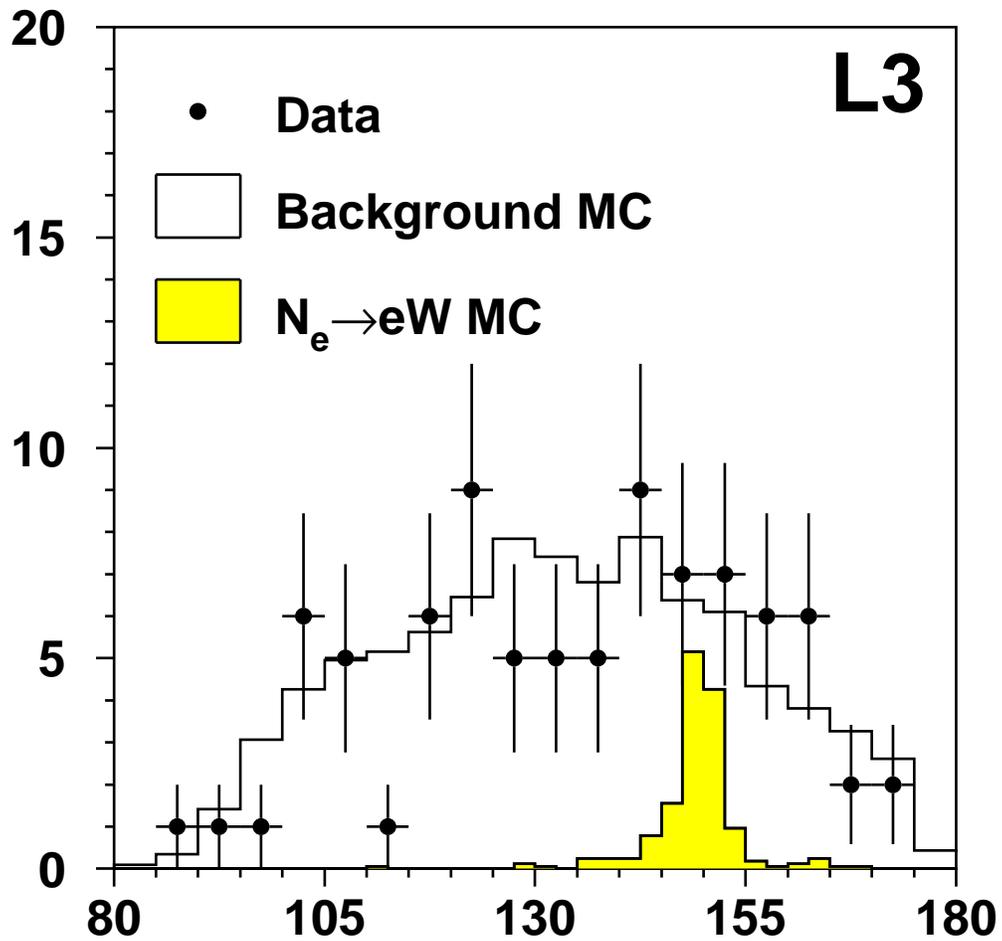}}
\end{center}
\caption
{
Distribution of the invariant mass of the event
after the kinematic fit. The points are
the data and the solid histogram is the background MC.
The shaded histogram is the predicted
signal e$^+ $e$^- \rightarrow \nu \mathrm{N}$ for a 150 \GeV\ heavy
neutrino. 
The normalization for the signal Monte Carlo is arbitrary. 
}
\label {antot}
\end{figure}

\begin{figure}[p]
\begin{center}
\mbox{\epsfysize=16cm\epsffile{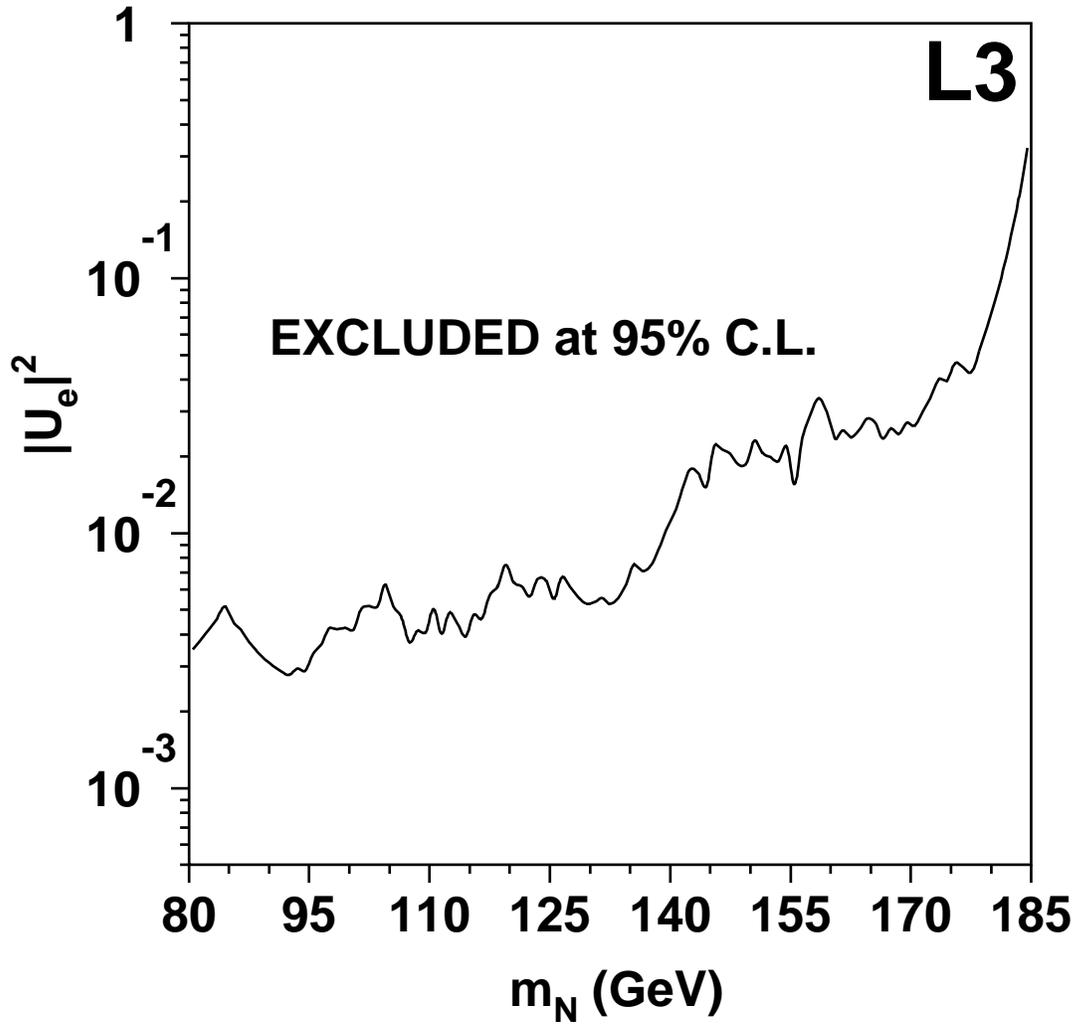}}
\end{center}
\caption
{
Upper limit at the 95\% C.L. on the mixing amplitude  as a function
of the singlet heavy neutrino mass.
}
\label {limit_maj}
\end{figure}

\end{document}